\documentclass[%
reprint,
superscriptaddress,
 amsmath,amssymb,
 aps,
prb,
]{revtex4-2}

\usepackage{graphicx}
\usepackage{dcolumn}
\usepackage{bm}
\usepackage{siunitx} 
\usepackage{braket}

\usepackage{xcolor}
\usepackage[normalem]{ulem}
\usepackage{soul}

\usepackage[pagebackref=false]{hyperref}

\begin{document}
\author{K.~Hecker}
\author{S. M\"oller}
\affiliation{JARA-FIT and 2nd Institute of Physics, RWTH Aachen University, 52074 Aachen, Germany,~EU}%
\affiliation{Peter Gr\"unberg Institute  (PGI-9), Forschungszentrum J\"ulich, 52425 J\"ulich,~Germany,~EU}
\author{T. Deußen}
\thanks{Present address: CST - Compound Semiconductor Technology, RWTH Aachen University, 52074 Aachen, Germany}
\affiliation{JARA-FIT and 2nd Institute of Physics, RWTH Aachen University, 52074 Aachen, Germany,~EU}%
\author{H. Dulisch}
\affiliation{JARA-FIT and 2nd Institute of Physics, RWTH Aachen University, 52074 Aachen, Germany,~EU}%
\affiliation{Peter Gr\"unberg Institute  (PGI-9), Forschungszentrum J\"ulich, 52425 J\"ulich,~Germany,~EU}
\author{L. Banszerus}
\thanks{Present address: University of Vienna, Faculty of Physics, Boltzmanngasse 5, 1090 Vienna, Austria}
\affiliation{JARA-FIT and 2nd Institute of Physics, RWTH Aachen University, 52074 Aachen, Germany,~EU}%
\affiliation{Peter Gr\"unberg Institute  (PGI-9), Forschungszentrum J\"ulich, 52425 J\"ulich,~Germany,~EU}
\author{K.~Watanabe}
\affiliation{Research Center for Electronic and Optical Materials, National Institute for Materials Science, 1-1 Namiki, Tsukuba 305-0044, Japan}
\author{T.~Taniguchi}
\affiliation{Research Center for Materials Nanoarchitectonics, National Institute for Materials Science, 1-1 Namiki, Tsukuba 305-0044, Japan}%

\author{C.~Volk}
\author{C.~Stampfer}
\email{stampfer@physik.rwth-aachen.de}
\affiliation{JARA-FIT and 2nd Institute of Physics, RWTH Aachen University, 52074 Aachen, Germany,~EU}%
\affiliation{Peter Gr\"unberg Institute  (PGI-9), Forschungszentrum J\"ulich, 52425 J\"ulich,~Germany,~EU}%

\title{Probing charge noise in bilayer graphene quantum dots by Landau-Zener-St\"uckelberg-Majorana spectroscopy}
\date{\today}

\begin{abstract}
Charge noise is an important factor limiting qubit coherence and relaxation in solid-state devices. In bilayer graphene (BLG) quantum dots, recently established as a promising platform for spin- and valley-based qubits, both the origin and magnitude of charge noise remain largely unexplored. Here, we investigate high-frequency charge noise using Landau-Zener-Stückelberg-Majorana (LZSM) interference spectroscopy. We study a single-particle charge qubit formed in a BLG double quantum dot at frequencies between 5 and $\SI{10}{GHz}$ and extract a noise spectral density $S_\varepsilon$ 
on the order of 0.5--$\SI{0.9}{neV}/\sqrt{\mathrm{Hz}}$. This is comparable to values reported for III-V semiconductor platforms and silicon. From the temperature and frequency dependence of the charge qubit decoherence, we conclude that thermal (Johnson) noise or electron-phonon coupling dominates over two-level fluctuators.

\end{abstract}
\maketitle
\section{Introduction}
Quantum-mechanical two-level systems are the fundamental building blocks of quantum computation and quantum simulation and are commonly referred to as qubits. When a two-level system is driven through an avoided crossing, it can undergo a transition between its energy levels, known as a Landau-Zener (LZ) transition~\cite{Landau1932, Zener1932Sep}. Coherent driving through repeated LZ transitions gives rise to Landau-Zener-Stückelberg-Majorana (LZSM) interference~\cite{stueckelberg1932theory,majorana1932atomi,Shevchenko2010Jul}. This phenomenon has been widely used to control and characterize qubits across a range of platforms, including superconducting qubits~\cite{Oliver2005Dec,Sillanpaa2006May,Wilson2010Jan} and charge or spin qubits realized in semiconductor quantum dots~\cite{Dupont-Ferrier2013Mar,Cao2013Jan,Petta2010Feb,Stehlik2012Sep,Forster2014Mar,Forster2015Dec,Chatterjee2018Jan,Mi2018Oct}. 
In particular, detailed studies of LZSM interference provide direct access to the relaxation and coherence times of a qubit, two important quantities that limit its performance and are often strongly affected by magnetic and electrical noise~\cite{Loss1998Jan,Petta2005Sep,Dial2013Apr,Yoneda2018Feb}.

Bilayer graphene (BLG) has emerged as a promising platform for spin- and valley-based qubits, as its intrinsically weak hyperfine interaction and spin-orbit coupling are expected to support long coherence times~\cite{Trauzettel2007Mar}. Its electrostatically tunable band gap~\cite{Zhang2009Jun,Icking2022Nov} enables the confinement of single charge carriers in quantum dots (QDs)~\cite{Banszerus2018Aug,Eich2018Aug} and the investigation of their spin and valley structure~\cite{Eich2018Jul,Banszerus2021Sep,Kurzmann2019Jul,Moller2021Dec}. Although relaxation times of single-particle spin and valley states have been studied as a function of their energy splitting from the ground state~\cite{Banszerus2022Jun,Gachter2022May,Banszerus2025Jul,Denisov2025Apr,Wang2024Jul,Wang2025Oct}, and coherent charge oscillations have been demonstrated~\cite{Hecker2023Nov}, many mechanisms underlying relaxation and decoherence in BLG QDs remain unresolved. Among them is the role of temperature-activated electrical noise, including charge noise, thermal noise, and electron-phonon coupling, is still largely unknown. These noise sources can couple directly to charge and indirectly to spin through spin-orbit interaction. Their quantitative impact, especially at high frequencies relevant for small level splittings, remains to be established.

\section{Results}

Here, we report high-frequency noise spectroscopy of a single-particle charge qubit hosted in a bilayer graphene double quantum dot (DQD). We extract the noise spectral density, $S_\varepsilon$, from temperature-dependent LZSM interference measurements in the GHz regime~\cite{Rudner2008Nov,Ivakhnenko2023Jan,Forster2014Mar} and obtain values in the range of 
$S_\varepsilon=0.5$--$\SI{0.9}{neV}/\mathrm{\sqrt{Hz}}$. Remarkably, despite the close proximity of the gate electrodes to the QDs and the different dielectric environment, these values are comparable to those reported for conventional semiconductor platforms, where $S_\varepsilon$ is typically 0.2 to $\SI{1}{neV}/\mathrm{\sqrt{Hz}}$~\cite{Dial2013Apr,Connors2022Feb}. From the temperature ($0.08$--$\SI{1.2}{K}$) and frequency ($4$--$\SI{10}{GHz}$) dependence of the 
noise spectral density, we identify
electron-phonon coupling or Johnson noise from nearby gate electrodes and contacts as the main source of decoherence.

\begin{figure}[h!]  
    \centering
    \includegraphics[draft=false,keepaspectratio=true,clip,width=\linewidth]{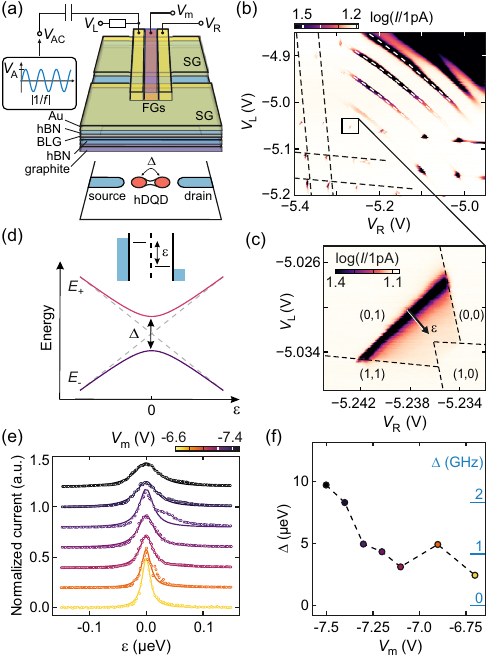}
    \caption{
    (a) Schematic illustration of the 2D heterostructure consisting of BLG encapsulated in hBN on a graphite gate. Cr/Au split gates (SGs) and two layers of finger gates (FGs) are deposited on top, separated by $\SI{15}{nm}$ of $\mathrm{Al_2O_3}$.
    Voltages applied to the SGs and FGs define an $n$-type channel and a hole DQD (hDQD) with tunable interdot tunnel coupling $\Delta$. 
    A microwave signal with frequency $f$
    can be applied to the left FG via a bias-tee.
    (b) Charge stability diagram showing the logarithmic current $I$ through the channel as a function of $V_\mathrm{R}$ and $V_\mathrm{L}$ at $V_\mathrm{SD}=\SI{700}{\mu V}$. The white dashed lines mark the regime where an electron QD forms between the left and right FGs, while at lower gate voltages a hole DQD is formed. 
    (c) Close-up of the charge transition $(0,1)\rightarrow(1,0)$, highlighted by the black box. The arrow indicates the detuning axis $\varepsilon$.
    (d) Energy spectrum of the single-particle charge qubit as a function of detuning $\varepsilon$, highlighting the  avoided crossing with energy splitting $\Delta$ at $\varepsilon=0$.
    (e) Line traces across the resonant line in plane (b) at $\varepsilon=0$ for different values of $V_\mathrm{m}$. Lorentzian line shapes (Eq.~(\ref{eq1})) are fitted to extract $\Delta$. The current is normalized, and the traces are vertically offset for clarity.
    (f) Tunnel coupling $\Delta$ as a function of $V_\mathrm{m}$, extracted from the fits in panel~(d).
    }
    \label{f1}
\end{figure}
To study the coherence of the charge degree of freedom in BLG, we use a DQD hosting a single charge carrier that can occupy either the left or the right QD, thereby forming a charge qubit~\cite{Petersson2010Dec,Hecker2023Nov}.
%
The DQD is realized in the device schematically depicted in Fig.~\ref{f1}(a). It consists of a heterostructure of BLG, encapsulated in hBN, placed on a graphitic back gate. 
Split gates (SGs) are fabricated on top to define a one-dimensional channel in the BLG, while three finger gates (FGs) across the channel allow the electrostatic confinement of the QDs.
For details on device fabrication, see Ref.~\cite{Banszerus2020Oct}.
The voltages $V_\mathrm{L}$ and $V_\mathrm{R}$ control the electrochemical potentials of the two QDs individually (see labels in Fig.~\ref{f1}(a)). 
The left FG is connected to a bias-tee to apply a microwave excitation ($V_\mathrm{AC}$) with frequency $f$ and amplitude $V_\mathrm{A}$ to perform LZSM spectroscopy measurements. 
An additional middle gate, at voltage $V_\mathrm{m}$, tunes the interdot tunnel coupling $\Delta$~\cite{Banszerus2021Mar}.
The charge stability diagram in Fig.~\ref{f1}(b) reveals different electrostatic regimes, ranging from 
a single electron QD formed  in between the left and right $\mathrm{FG}$ (white dashed lines) to the formation of a hole-type DQD (lower left). 
Considerations of the electrostatic environment are discussed in detail in Ref.~\cite{Banszerus2020Mar}. The black box highlights the first charge transition between the left and right hole QD, shown as a close-up in Fig.~\ref{f1}(c).
We define the detuning axis $\varepsilon$ along which the QD levels are shifted relative to each other.
Resonant transport arises at $\varepsilon=0$, where the ground states of the QDs are aligned. 
For $\varepsilon>0$, transport through the DQD is substantially suppressed, reflecting a low interdot coupling, $\Delta$~\cite{Banszerus2021Mar}. 
Figure~\ref{f1}(d) shows the energy spectrum of the two-level system as a function of detuning $\varepsilon$, highlighting the avoided crossing arising from the coupling between the two QDs. The double quantum dot, occupied by a single charge carrier, is described by the Hamiltonian
$H= -\Delta/2 \sigma_x - \varepsilon/2\sigma_z$, where $\sigma_i$ ($i=x,y,z$) are the Pauli matrices. 
In this basis, the charge carrier can occupy either the left or the right quantum dot, $\ket{L}$ or $\ket{R}$, such that the system realizes a charge qubit~\cite{Hayashi2003Nov,Gorman2005Aug,Petersson2010Dec,Stehlik2012Sep,Cao2013Jan,Dovzhenko2011Oct}.
In the low coupling regime ($\Gamma_\mathrm{L},\Gamma_\mathrm{R} \gg \Delta/2h$), transport through the DQD can be described by~\cite{Stoof1996Jan,vanderWiel2002Dec}
\begin{equation}
    I(\varepsilon) = e/h\frac{\Delta^2/h\Gamma_\mathrm{R}}{1+(2\varepsilon/h\Gamma_\mathrm{R})^2},
\label{eq1}
\end{equation}
where $\Gamma_\mathrm{L}$ and $\Gamma_\mathrm{R}$ are the tunnel rate to the left and right lead, respectively.
Fig.~\ref{f1}(e) shows line cuts along the detuning axis as function of the middle gate voltage, $V_\mathrm{m}$ and the corresponding fit of Eq.~(\ref{eq1}). 
The fit results are depicted in Fig.~\ref{f1}(f) and show a strong dependence of $\Delta$ on $V_\mathrm{m}$ with values of $\Delta \approx 2.4-9.7~ \text{$\mu$eV}$, corresponding to $0.6-2.3~\text{GHz}$.

To investigate the coherence of this charge qubit, we employ LZSM interference measurements. A TLS driven across the avoided crossing at $\varepsilon=0$ undergoes a LZ transition from the ground state to the excited state with probability $P_\mathrm{LZ} = e^{-\pi\Delta^2/hv}$, where $v$ is the sweep velocity~\cite{Shevchenko2010Jul}.
Furthermore, the transition will introduce a relative Stokes phase $\phi$~\cite{Ivakhnenko2023Jan}. Interference arises from subsequent passages through the avoided crossing, with a phase difference 
that depends on the energy splitting and the elapsed time between the transitions.

\begin{figure*}[th]  
    \centering
    \includegraphics[draft=false,keepaspectratio=true,clip,width=\linewidth]{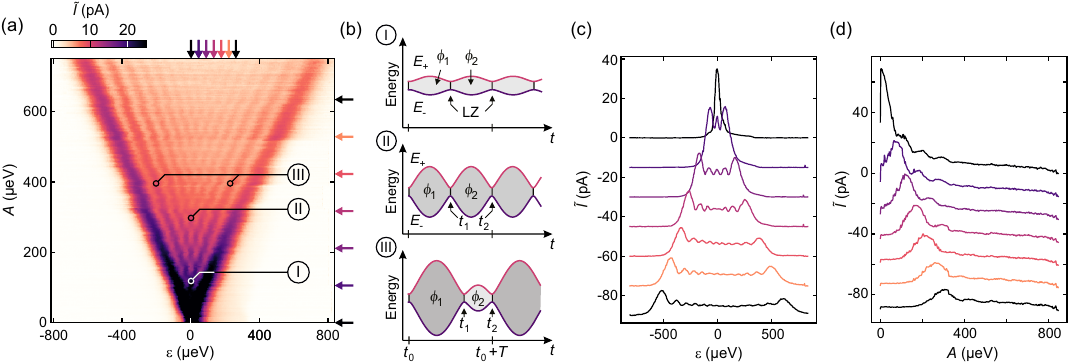}
    \caption{
    (a) LZSM interference pattern showing the current as a function of $\varepsilon$ and the microwave amplitude $A$. The measurements were taken at a frequency of $f=\SI{6.6}{GHz}$ and a base temperature of $T=\SI{90}{mK}$ with $V_\mathrm{m}=\SI{-7.4}{V}$ and $V_\mathrm{SD}=\SI{1}{mV}$. 
    A line-by-line background has been subtracted from the current $I$ to improve the visibility of the pattern; the resulting quantity is denoted throughout the manuscript by $\tilde{I}$.
    (b) Schematic representations of the time evolution of the states $E_+$ and $E_-$ (c.f. Fig.~\ref{f1}(d)) for different combinations of $A$ and $\varepsilon$ (I$-$III). While the total duration of one oscillation period $T=1/f$, remains constant, the relative timing of the LZ transitions ($t_1$, $t_2$) and the phase accumulated between them ($\phi_1$, $\phi_2$) vary with $\varepsilon$ and $A$.
    Schemes I and II differ only in amplitude, with $\varepsilon$ fixed at 0, whereas in scheme III the detuning is finite (for more details see text).
    (c) Line cuts along the detuning axis of the measurement shown in panel (a) taken at different values of $A$ (horizontal arrows in panel (a)). 
    (d) Line cuts along the amplitude axis of the measurement in panel (a) at different values of $\varepsilon$ (vertical arrows in panel (a)). For a better visibility, an offset of multiples of $\SI{10}{pA}$ has been subtracted from the line traces in (c) and (d).
    }
    \label{f2}
\end{figure*}
We measure this interference, by applying a continuous microwave voltage signal $V_\mathrm{AC}(t)=V_\mathrm{A} \sin{(2\pi f t)}$ to the left FG with amplitude $V_\mathrm{A}$ and frequency $f$. 
Fig.~\ref{f2}(a) displays the resulting LZSM interference pattern, 
showing oscillations of the current as a function of the effective amplitude $A$ and the detuning energy~\cite{Shevchenko2010Jul,Forster2014Mar} when applying a frequency of $f=\SI{6.6}{GHz}$.
The effective amplitude $A=\alpha_\mathrm{AC}V_\mathrm{A}$ 
at the device includes the effects of cryogenic attenuation, frequency-dependent line losses, and the gate lever arm.
When applying the sinusoidal voltage pulse, a maximum of two LZ transitions can appear during one oscillation period $T=1/f$.
The condition of the interference depends on the phase differences ($\phi_1$, $\phi_2$) of the wave functions accumulated between these two subsequent LZ transitions~\cite{Shevchenko2010Jul}
\begin{equation}
\phi_\mathrm{i} =\frac{1}{2\hbar} \int_{t_\mathrm{i-1}}^{t_\mathrm{i}} \sqrt{\varepsilon(t)^2+\Delta^2} \,dt = \frac{1}{2\hbar}\int_{t_\mathrm{i-1}}^{t_\mathrm{i}} (E_\mathrm{+}-E_-)dt,
\end{equation}
with $i = 1,2$. The schematic in Fig.~\ref{f2}(b) (regime I) illustrates the eigenstates ($E_+,\ E_-$) of the DQD as they evolve in time, while a continuous sine voltage is applied symmetrically around the avoided crossing ($\varepsilon=0$). 
The enclosed areas between subsequent LZ transitions (gray shaded areas) are proportional to $\phi_\mathrm{1}$ and $\phi_\mathrm{2}$. When increasing the amplitude $A$ (compare regimes I and II), both $\phi_\mathrm{1}$ and $\phi_2$ grow simultaneously. 
When increasing the detuning energy $\varepsilon$ (compare regimes II and III), the ratio between the two enclosed areas during one oscillation period is changing. 
The final state after two successive transitions becomes the initial state for the next drive period and, depending on the interference conditions, can lead to an iterative change in the occupation probability of the excited state.
Under resonance condition, this leads to Rabi-like oscillations with a time-average solution~\cite{Ivakhnenko2023Jan}. Considering that the measurement is predominantly performed in the diabatic (fast passage) limit ($A\gg \Delta^2/hf$), the resonance condition is $\phi_1-\phi_2=k\pi$~\cite{Ivakhnenko2023Jan}\footnote{Note, that in this reference, they refer to $\phi_1$ as $\zeta_1$ and $\phi_2$ as $\zeta_2$.}. When the TLS is coupled to a dissipative environment by taking relaxation and decoherence rates into account, the interference pattern in Fig.~\ref{f2}(a) can be described in a rotating-wave approximation (see Ref.~\cite{Ivakhnenko2023Jan}).

Line cuts along both axes of the data in Fig.~\ref{f2}(a), taken at the positions marked by the horizontal and vertical arrows, are shown in Figs.~\ref{f2}(c) and~\ref{f2}(d).
As expected, they exhibit oscillations in the current along both the amplitude and detuning axes, which decay exponentially due to the loss of phase information (i.e. loss of coherence) through coupling to the dissipative environment.
\begin{figure*}[t]  
    \centering
    \includegraphics[draft=false,keepaspectratio=true,clip,width=\textwidth]{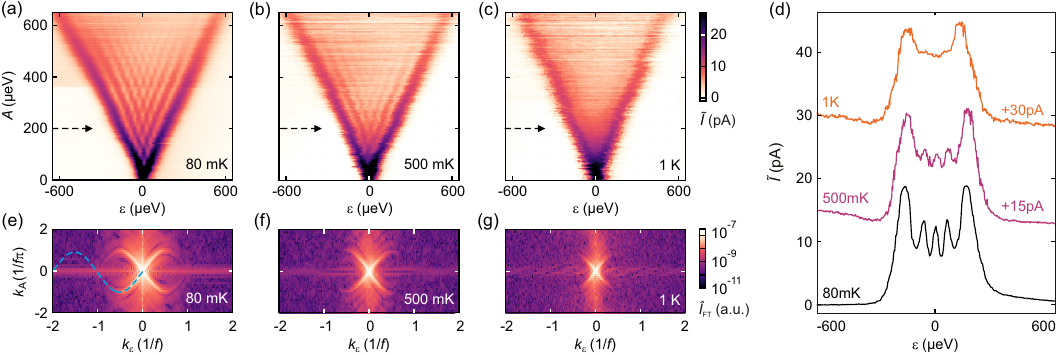}
    \caption{(a-c) LZSM interference patterns, as in Fig~\ref{f2}(a), measured at $T=\SI{80}{mK}$ (a), $T=\SI{500}{mK}$ (b) and $T=\SI{1}{K}$ (c), respectively. 
    (d) Line cuts along the black dashed lines of the interference patterns. For a better visibility, an offset of multiples of $\SI{15}{pA}$ has been added.
    (e-g) 2D Fourier transformation of the current $\hat{I}_\mathrm{FT}$ of the corresponding measurements shown in panels (a-c). The frequencies $k_\mathrm{A}$ and $k_\varepsilon$ have been normalized to the driving frequency $f=\SI{6.6}{GHz}$. The blue dashed line in panel (e) marks one branch of the sinusoidal arc described by Eq.~(\ref{eq4}).}   
    \label{f3}
\end{figure*}
To study the decoherence of the charge state systematically, we use quantum phase tomography via 2D Fourier transformation of the LZSM interferograms~\cite{Rudner2008Nov}: 
\begin{equation}
    I_\mathrm{FT}=\int\int_{-\infty}^\infty \exp(-ik_\mathrm{\varepsilon}\varepsilon-ik_\mathrm{A}A)I(\varepsilon,A)\mathrm{d\varepsilon}\mathrm{d}A.
\end{equation}
Figs.~\ref{f3}(a) and~\ref{f3}(e) show an LZSM interferogram and its corresponding discrete 2D Fourier transform. The axes $k_\varepsilon$ and $k_\mathrm{A}$ are normalized to the applied frequency of $f=\SI{6.6}{GHz}$. Apart from straight lines along $k_\varepsilon=0$ and $k_\mathrm{A}=0$ in Fig.~\ref{f3}(e), which are an artifact from the discrete Fourier transformation~\cite{Ivakhnenko2023Jan}, a sinusoidal shape is appearing in the center. The condition for these sinusoidal arcs is generally given by~\cite{Rudner2008Nov,Ivakhnenko2023Jan,Forster2014Mar}:
\begin{equation}\label{eq4}
    k_\mathrm{A}(k_\varepsilon)=\pm \frac{2 l}{2\pi f}\sin \left({\frac{2\pi f\, k_\varepsilon+2\pi l'}{2l}}\right),
\end{equation}
where $l'=0,1,2,...$ and $l=1,2,3,...$, while $l'<l$. In Fig.~\ref{f3}(e), only the main sinusoidal ($l'=0$, $l=1$) is visible. 
A qualitative comparison with the numerical calculations in Ref.~\cite{Ivakhnenko2023Jan} suggests that the relaxation rate $\Gamma_1$ is of the same order as the time interval between the two LZ transitions. In contrast, the disappearance of the sinusoidal modulation along the $k_\varepsilon$ axis is governed by the decoherence rate $\Gamma_2$.
When increasing the temperature, the LZSM interference pattern disappears, as can be seen in Fig.~3(b) and (c). Line cuts along the $k_\varepsilon$-axis at $A=\SI{200}{\mu eV}$ in Fig.~\ref{f3}(d) show that the oscillation maxima disappear as function of the temperature. 
The disappearance of the LZSM interference is also visible in the Fourier transformation depicted in Fig.~\ref{f3}(e-g), where the arcs of the sinusoidal vanish with temperature.

\begin{figure*}[t]  
    \centering
    \includegraphics[draft=false,keepaspectratio=true,clip,width=1.0\linewidth]{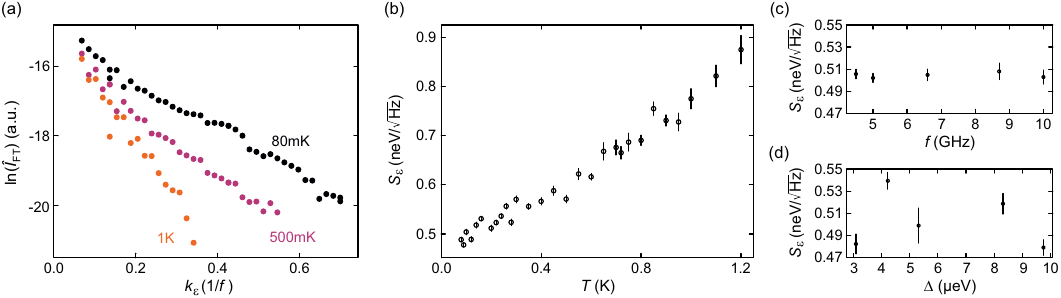}
    \caption{
    (a) Logarithmic amplitude of the current $\hat{I}_\mathrm{FT}$ along the sinusoidal arcs (shown in Figs.~\ref{f3}(e),~\ref{f3}(f) and~\ref{f3}(g)) as function of $k_\varepsilon$ for three different temperatures.
    (b) Noise spectral density $S_\varepsilon$ as function of  temperature $T$. (c) $S_\varepsilon$ as function of the applied frequency $f$.  (d) $S_\varepsilon$ as function and the interdot coupling $\Delta$, which is tuned by an appropriate voltage applied to the middle gate, as shown in Fig.~\ref{f1}(f).
    }
    \label{f4}
\end{figure*}

To quantitatively assess the decoherence, we evaluate the amplitude of the sinusoidal following Eq.~(\ref{eq4}) with $l'=0$ and $l=1$. Along the arcs, the amplitude is decaying with~\cite{Rudner2008Nov,Ivakhnenko2023Jan,Forster2014Mar,Cao2013Jan,Gonzalez-Zalba2016Mar}
\begin{equation}\label{eq6}
    \hat{I}_\mathrm{FT}(k_\varepsilon(k_\mathrm{A}))\propto e^{-\lambda|k_\varepsilon|}+e^{-\lambda^*k_\varepsilon^2/2}.
\end{equation}
Fig.~\ref{f4}(a) shows the extracted amplitudes $\hat{I}_\mathrm{FT}(k_\varepsilon
(k_\mathrm{A}))$, exemplary for base temperatures of $T=\SI{80}{mK},\ \SI{500}{mK}$ and $\SI{1}{K}$, in a logarithmic scale. We observe a linear dependence and therefore extract the slope $\lambda$ (Eq.~(\ref{eq6})). 
The fit parameter $\lambda$ is a measure of the decoherence rate $\Gamma_2$~\cite{Ivakhnenko2023Jan,Rudner2008Nov}. 
In the Markovian limit of uncorrelated noise, this parameter is related to the noise spectral density by $S_\varepsilon=\sqrt{2\hbar\lambda}$.
As shown in Fig.~\ref{f4}(b), the noise spectral density exhibits a pronounced temperature dependence, consistent with the disappearance of the LZSM interference pattern in Fig.~\ref{f3}. By contrast, no clear dependence is found on either the driving frequency in the range $f\approx4$--$\SI{10}{GHz}$ (Fig.~\ref{f4}(c)), or the interdot coupling~$\Delta$ (Fig.~\ref{f4}(d)).
\section{Discussion}
Possible sources of decoherence, all of which are expected to yield a noise spectral density that increases with temperature, include (i) charge noise arising from statistical two-level fluctuators (TLFs) in the electrostatic environment~\cite{Buizert2008Nov,Connors2019Oct,Gungordu2019Feb}, (ii) thermal noise, or Johnson-Nyquist noise, coupling into the system through gate electrodes and contacts~\cite{Connors2022Feb}, and (iii) electron-phonon coupling between the charge degree of freedom and the phonon bath of the material~\cite{Hayashi2003Nov,Petit2018Aug}.
The influence of thermally activated TLFs has also been studied in silicon QDs~\cite{Connors2019Oct,Petit2018Aug}, where a linear temperature dependence of the decoherence rate was observed.
In both silicon and III-V semiconductor systems, charge noise arising from TLFs typically dominates at low frequencies and approximately follows a $1/f$ dependence.
At higher frequencies, this contribution is usually overtaken by a lower-amplitude white-noise background~\cite{Dial2013Apr,Connors2019Oct,Kenyon2000Dec,Connors2022Feb,Basset2014Aug,Buizert2008Nov,Fujisawa2000Jul,Petit2018Aug}.
Since our measurements show a frequency-independent noise spectral density of $S_\varepsilon\approx0.5~\mathrm{neV/\sqrt{Hz}}$ over the frequency range from 4 to $\SI{10}{GHz}$ (see Fig.~\ref{f4}(c)), 
we can rule out TLFs as the dominant source of decoherence in this regime. We can also exclude quantum noise (Nyquist noise), which is expected to show a pronounced frequency dependence~\cite{Astafiev2004Dec}.
This leaves thermal noise (Johnson noise) as a relevant contribution, which can couple into the system through the microwave source and coaxial cables or arise from electric-field fluctuations within the bilayer-graphene-based van der Waals heterostructure itself. Such contributions appear as white noise and can become the dominant limitation at high frequencies, where $1/f$-noise is strongly  suppressed~\cite{Connors2022Feb,Petit2018Aug}.
In addition, electron-phonon coupling provides a plausible explanation for the loss of coherence at high frequencies and exhibits a temperature dependence consistent with our measurements~\cite{Hayashi2003Nov,Forster2014Mar}. We therefore conclude that Johnson noise and electron-phonon coupling remain the most likely sources of decoherence in our device. These findings help narrow down the microscopic origin of charge noise in bilayer graphene quantum dots and provide an important basis for optimizing future spin- and valley-qubit devices in this material platform.
\newline
\newline

\textbf{Acknowledgements}
The authors thank F.~Hassler for fruitful discussions,  F.~Lentz, S.~Trellenkamp and M.~Otto for help with sample fabrication.
This project has received funding from the European Research Council (ERC) under grant agreement No. 820254, the Deutsche Forschungsgemeinschaft (DFG, German Research Foundation) under Germany's Excellence Strategy - Cluster of Excellence Matter and Light for Quantum Computing (ML4Q) 2004/2 - 390534769, through the DFG SPP 2244 (Project No. 535377524), the FLAG-ERA grant ThinQ, by the DFG - 534269806, and by the Helmholtz Nano Facility~\cite{Albrecht2017May}. 
K.W. and T.T. acknowledge support from the JSPS KAKENHI (Grant Numbers 21H05233 and 23H02052), the CREST (JPMJCR24A5), JST and World Premier International Research Center Initiative (WPI), MEXT, Japan.\\

\textbf{Author contributions  }
K.H., S.M., C.V. and C.S. conceived this experiment.
K.H, L.B. and S.M. fabricated the device. K.H., S.M. and H.D. performed the measurements and analyzed the data with the help of T.D.. K.W. and T.T. synthesized the hBN crystals. C.V. and C.S. supervised the project. K.H., S.M., H.D., C.V., and C.S. wrote the manuscript with contributions from all authors.\\

\textbf{Data availability}
The data underlying this study are openly available in a Zenodo repository at DOI 10.5281/zenodo.20137610.\\

\textbf{Competing interests  }
The authors declare no competing interests.

%

\end{document}